\documentclass[aps,prb,twocolumn,showpacs,preprintnumbers,amsmath,amssymb,superscriptaddress]{revtex4-1}

\usepackage[dvips]{graphicx}
\usepackage{dcolumn}
\usepackage{bm}
\usepackage{color}
\usepackage{amsmath}
\usepackage{amsbsy}
\usepackage[utf8]{inputenc}
\newcommand{\e}{\varepsilon}
\newcommand{\be}{\begin{equation}}
\newcommand{\ee}{\end{equation}}
\newcommand{\bea}{\begin{eqnarray}}
\newcommand{\eea}{\end{eqnarray}}
\def \ket #1{\left| #1 \right\rangle}

\def \scalarprod #1#2{\left \langle #1 \right. \left| #2 \right\rangle}

\begin{document}
 
\title{Anderson Localization and Quantum Hall Effect: Numerical Observation of Two Parameter
Scaling}

\author{Miklós Antal Werner}
\affiliation{MTA-BME Exotic Quantum Phases ``Momentum'' Research Group\\
Department of Theoretical Physics, Budapest University of Technology and Economics, 1111 Budapest, Budafoki út 8, Hungary}

\author{Arne Brataas}
\affiliation{Department of Physics, Norwegian University of Science and Technology, NO-7491 Trondheim, Norway}

\author{Felix von Oppen}
\affiliation{\mbox{Dahlem Center for Complex Quantum Systems and Fachbereich Physik, Freie Universität Berlin, 14195 Berlin, Germany }}

\author{Gergely Zaránd}
\affiliation{MTA-BME Exotic Quantum Phases ``Momentum'' Research Group\\
Department of Theoretical Physics, Budapest University of Technology and Economics, 1111 Budapest, Budafoki út 8, Hungary}

\date{\today}

\begin{abstract}
A two dimensional disordered system of non-interacting fermions in a homogeneous magnetic field is investigated numerically. 
By introducing a new magnetic gauge, we explore the renormalization group (RG) flow of the longitudinal and Hall  
conductances with higher precision than previously studied, and find that the flow is consistent with
the predictions of Pruisken and Khmelnitskii.
The extracted critical exponents agree with the results obtained by using transfer matrix methods.
The necessity of a second parameter is also reflected in the level curvature distribution. Near the critical point the 
distribution slightly differs from the prediciton of random matrix theory, in agreement with previous works. Close to the quantum Hall
fixed points the distribution is lognormal since here states are strongly localized.

\end{abstract}


\maketitle

\section{Introduction}
Topological phases of quantum systems have been at the focus of intense studies in recent years.\cite{HasanKane} Many topological insulators are exotic 
band insulators where the energy bands are characterized by non-trivial topological quantum numbers. 
These topological quantum numbers reflect the non-trivial topological ground state structure, 
arising from the symmetries and the dimensionality of the system.\cite{TopInsPerTable}
In a finite sample, the non-trivial topological structure of the ground state gives rise to topologically protected 
gapless edge states in the otherwise gapped system.
These edge states are protected by topology and are 
robust against perturbations and disorder which do not break the underlying symmetries of the system.

One of the simplest examples of an insulating state with a non-trivial topological structure is provided by the Integer Quantum Hall (QH) Effect.\cite{vonKlitzing} 
In a two dimensional electron gas, a homogeneous magnetic field splits the energy spectrum into Landau levels that are broadened into Landau bands by disorder. 
States within a Landau band are localized, except for a single critical, extended state at the center of each band.\cite{QHallDisorder}
Each Landau band is characterized by a non-trivial topological invariant, the 
Chern number.\cite{ThoulessChern} It can be shown that the Chern number of a band -- apart from a universal pre\-fac\-tor $e^2/h$ -- equals the contribution of the 
band to the Hall conductance. As a result, if the Fermi energy lies between two Landau bands, then the Hall conductance is the sum of the 
Chern numbers associated with the filled Landau bands. The topological character of this insulating phase is also manifested through the emergence of chiral
edge states:\cite{edge_exp1,edge_exp2,edge_exp3,edge_exp4} In fact, the total Chern number equals the number of chiral states.

 As mentioned above, in each Landau band there is a single delocalized state and an associated critical energy ($E_{c,i}$ in the $i$-th band). 
 Topologically distinct QH phases are separated by these critical states,\cite{Halperin, QHallCriticalEnergies}  
and near them a critical behavior is observed, with the localization length (the size of the localized wave functions) diverging as
\be
\xi \sim \left|{E-E_{c,i}}\right|^{-\nu} \; \mathrm{.}
\ee
Experimentally, a quantized Hall conductance is observed if the system size (or the inelastic scattering length, $L_{\rm{in}}$) is much larger than the localization
length at the Fermi energy, $\xi(E_{F})$. \cite{footnote1}

\begin{figure}
 \includegraphics[width = 0.35 \textwidth]{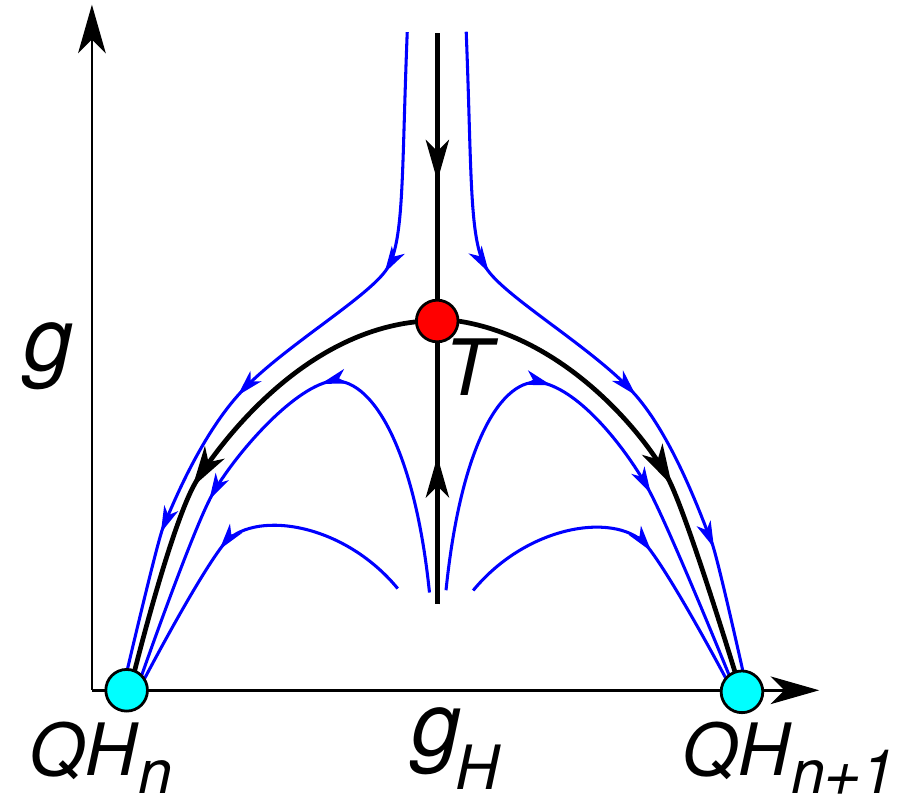}
 \caption{\textit{(Color online)} Sketch of the proposed two parameter renormalization flow, in terms of the diagonal and Hall conductances $g$ and $g_H$, respectively. The attractive
 QH fixed points at integer dimensionless Hall conductances and vanishing diagonal conductance are denoted by cyan circles ($QH_{n}$ and $QH_{n+1}$). The transition
 fixed point ($T$) at a finite Hall and diagonal conductance is denoted by the red circle.}\label{flowqualitatively}
\end{figure}

The characterization of the topological quantum phase transition at these critical energies was a challenging task. Based on nonlinear $\sigma$-model
calculations, Pruisken and Khmelnitskii proposed
a two parameter scaling theory, formulated in terms of the diagonal and offdiagonal elements of the dimensionless conductance 
tensor $g \equiv g_{xx}$ and $g_H \equiv g_{xy}$, respectively.\cite{Pruisken,Khmelnitskii} According to this theory, by
increasing the system size $L$ (or lowering the temperature), the conductances follow the 
trajectories of a two dimensional flow diagram (see Fig. \ref{flowqualitatively}),
\be \label{renormflow_eq}
\frac{d \ln g}{d \ln L} = \beta(g,g_H) \, ; \quad \frac{d \ln g_H}{d \ln L} = \beta_H(g,g_H) \, ,
\ee
determined by the universal beta functions $\beta(g,g_H)$ and $\beta_H(g,g_H)$. In this flow, attractive QH fixed points appear at integer dimensionless Hall conductances
and vanishing diagonal conductance. Each of these fixed points corresponds to a QH phase and is associated with a plateau in the Hall conductance. 
Between these attractive fixed points, other, hyperbolic fixed points emerge: these correspond to the critical state and describe transition between the QH plateaus.

While certain predictions of the universal scaling theory were confirmed experimentally, apart from some recent results for graphene,\cite{graphene_flow}
a direct numerical verification of the two parameter renormalization flow for ordinary QH systems has never been established.
In this paper, we demonstrate numerically  the two parameter scaling theory, and estimate 
the relevant and irrelevant critical exponents. 

To this end, we investigate a system of noninteracting, spinless, charged fermions on a square lattice, as described by the Hamiltonian 
\be \label{hamiltonian}
H = \sum_{i} \e_i c_i^\dag c_i - \sum_{\langle i,j \rangle} t_{ij} c_i^\dag c_j + h.c. \; \textrm{.}
\ee
Here $c_i^\dag$ and $c_i$ denote fermionic operators that create or annihilate a fermion on the lattice site $i$, respectively. The site energies $\e_i \in [-W/2, W/2]$ are uniformly and
independently distributed and the external magnetic field
is introduced by using the usual Peierls substitution\cite{Peierls} 
\be \label{eq_peierls}
t_{ij} =  e^{i 2\pi A_{ij}} \; ,
\ee
 with the lattice vector potential defined as
\be
A_{ij} = \frac{e}{h} \int_{i}^j \mathbf{A} \cdot d \mathbf{l} \; .
\ee

In this work, we introduce a new lattice gauge, which -- in contrast to the Landau gauge -- allows us to perform computations for small magnetic fields 
 corresponding to a \emph{single} flux quantum through the system. We then perform 
exact diagonalization at various system sizes, $L$, while applying twisted boundary conditions with phases $\phi_x$ and $\phi_y$ in the $x$ and $y$
directions, respectively. By studying the sensitivity of the energy levels $E_\alpha = E_\alpha(\underline{\phi})$ and eigenstates $\ket{\alpha} = \ket{\alpha(\underline{\phi})}$ to
the phase $\underline{\phi} = (\phi_x, \phi_y)$, we are able to determine $g(L)$ and $g_H(L)$, and reconstruct the renormalization group flow, Eq. (\ref{renormflow_eq}).
We indeed find that, as predicted by Pruisken and Khmelnitskii, the flow exhibits stable QH fixed points with quantized values of $g_H$ and $g = 0$.  
Neighboring QH fixed points are
separated by a critical point of a finite Hall  and diagonal conductance. The critical exponents extracted from the flow are in agreement with previous transfer matrix
results.\cite{QHallExponent_Slevin}

\subsection{Thouless formula and Hall conductance}
\begin{figure}
 \includegraphics[width = 0.3 \textwidth]{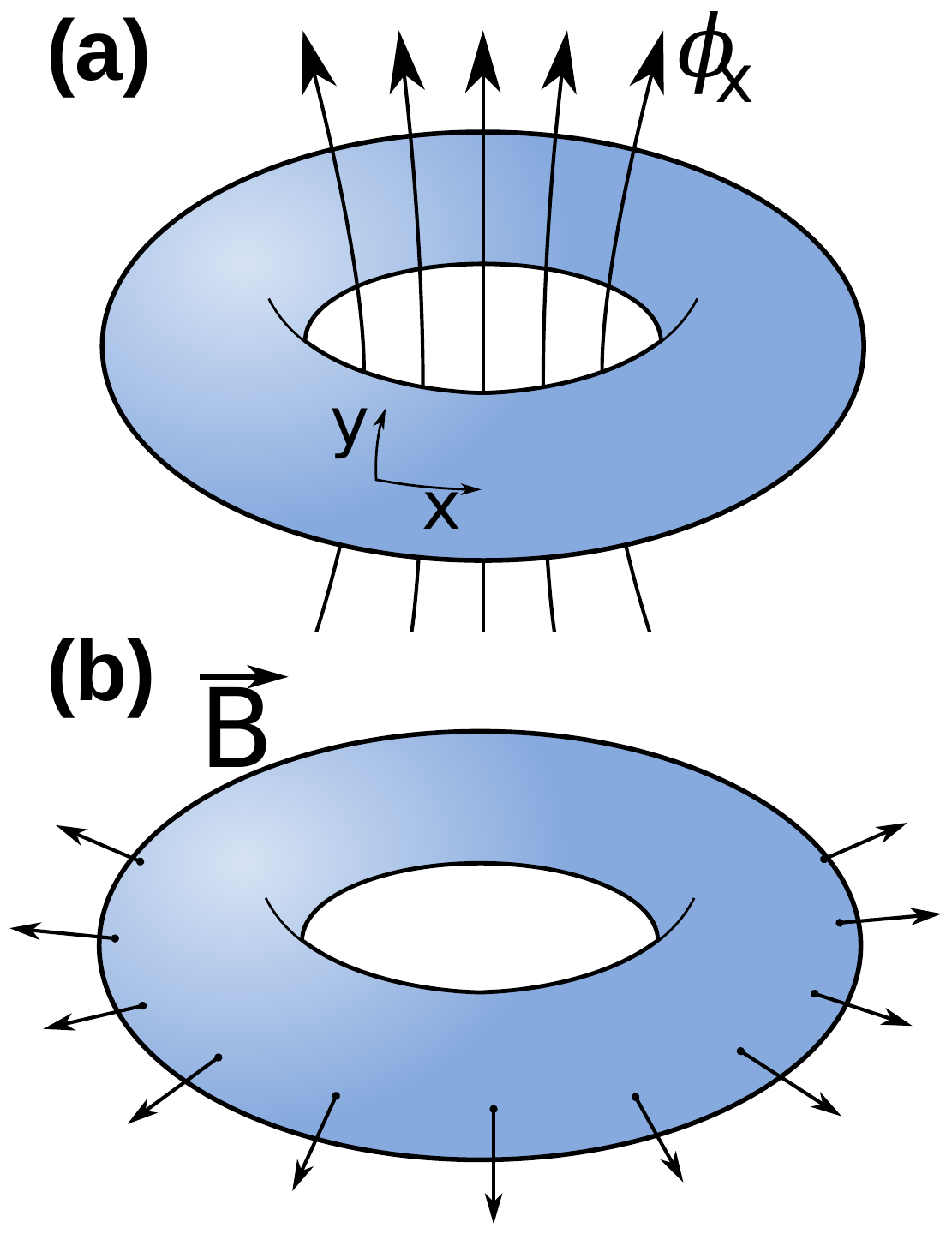}
 \caption{\textit{(Color online)} \textbf{(a)} The phase $\phi_x$ can be interpreted as a magnetic flux pierced through the torus. \textbf{(b)} 
 The external magnetic field
 pierces through the surface of the torus.}\label{torus} 
\end{figure}

The Kubo-Greenwood conductance formula\cite{KuboGreenwood} cannot be straightforwardly applied to a finite size system to extract its $T = 0$ temperature conductance in the 
thermodynamic limit. Fortunately, however, 
the Hall  and the diagonal conductances can both be related to the sensitivity of the states to the boundary conditions. The single particle eigenstates of Eq. (\ref{hamiltonian}) can 
be expanded as 
\be
\ket{\alpha} = \sum_{i} \alpha(i) \; c_i^\dag \ket{0} \; .
\ee
Labeling for the moment each site $i$ by its coordinates $i \rightarrow x,y$, a twisted boundary condition is defined by wrapping the system on a torus with the periodicity 
condition
\be 
\alpha(x + nL,y + mL) = e^{i(n \phi_x + m \phi_y)} \alpha(x,y) \; .
\ee
The phases $(\phi_x, \phi_y) = \underline{\phi}$ can be interpreted as magnetic fluxes pierced through the torus (and in its interior), while the external magnetic field
pierces through the surface of the torus (see Fig. \ref{torus}).
Solving the eigenvalue equation $H \ket{\alpha} = E_\alpha \ket{\alpha}$, one obtains the phase dependent eigenstates and eigenvalues 
$\ket{\alpha(\underline{\phi})}$ and $E_\alpha(\underline{\phi})$.

In a seminal work, Thouless and Edwards conjectured a relation between the diagonal conductance and the mean absolute curvature of 
eigenenergies at the Fermi energy,\cite{ThoulessFormula}
\bea \label{thouless_formula}
g \approx g_T & =&  \overline{|c_T(\alpha)|}_{E_\alpha = E_F} \; \textrm{,} \nonumber \\
 c_{T}(\alpha) & =& \frac{\pi}{\Delta(E_\alpha)} \frac{\partial^2 E_\alpha}{\partial \phi_x^2} \; \textrm{,}
\eea
with $\Delta(E_F)$ denoting the mean level spacing at the Fermi energy, and the overline indicating disorder averaging. 
Although this formula cannot be derived rigorously, it has been
verified numerically for a wide range of disorder.\cite{Montambaux} 

The Hall conductance can be directly related to the phase dependence of the eigenstates.\cite{HallKubo} In a finite system, 
the average Hall conductance at $T=0$ is
\be \label{Hall_conductance}
 g_{H} = \overline{\sum_{E_\alpha < E_F} c_{H}(\alpha)} \, .
\ee
Here $c_H(\alpha)$ denotes the Hall conductance of level $\ket{\alpha}$, and is given by
\be
c_{H}(\alpha) = 2 \pi i \left( \scalarprod{\frac{\partial \alpha}{\partial \phi_y}}{ \frac{\partial \alpha}{\partial \phi_x}} -
\scalarprod{\frac{\partial \alpha}{\partial \phi_x}}{ \frac{\partial \alpha}{\partial \phi_y}} \right) \; ,
\ee
the Berry curvature associated with $\ket{\alpha(\phi_x,\phi_y)}$.
In the following, we shall use Eqs. (\ref{thouless_formula}) and (\ref{Hall_conductance}) to determine the dimensionless conductances and establish the flow.

\subsection{Lattice gauge for small magnetic fields}
In a finite size system with periodic or twisted boundary condition, a homogeneous magnetic field cannot be arbitrary; 
the hopping matrix elements must respect the periodicity of the system, i.e., the hoppings at sites $(x + L,y)$ and $(x,y + L)$ must be equal with the one at $(x,y)$.
\begin{figure}
 \includegraphics[width = 0.3 \textwidth]{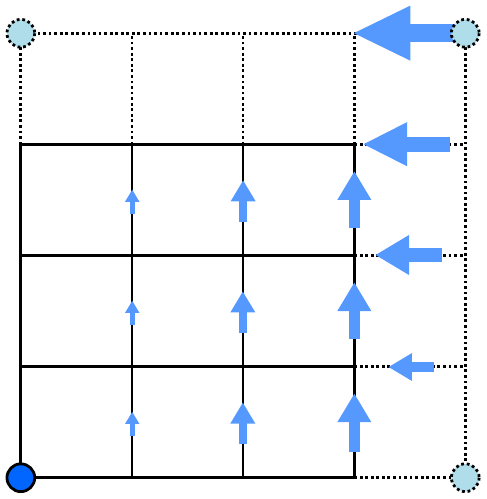}
 \caption{\textit{(Color online)} Sketch of bond vector potentials on a $4 \times 4$ lattice with periodic boundary condition: The circles denote equivalent sites.}\label{gauge} 
\end{figure}
The complex phases of the hopping matrix elements are related to the magnetic vector potential through the Peierls substitution, Eq. (\ref{eq_peierls}).

The periodicity of the system requires the complex phase of the hopping to be changed by $2 \pi \, n$ as the $x$ or $y$ coordinates are shifted by $L$, and imposes
restrictions on the total field pierced through the system.

The magnetic flux through a unit cell can be determined by summing the hopping phases around the cell, while the magnetic field in a cell can be defined as the flux divided by the area
of the cell. Setting the lattice size to $a = 1$, the magnetic field reads 
\bea
B_{(x+1/2\, ,\; y+1/2)} &=& \frac{h}{e} \left[ A_{(x,y)(x+1,y)} + A_{(x+1,y)(x+1,y+1)} + \right. \nonumber \\ &+& \left. A_{(x+1,y+1)(x,y+1)} + A_{(x,y+1)(x,y)} \right] \; . 
\nonumber \\ & &
\eea
Periodic boundary conditions imply that the total magnetic flux through the whole system is a multiple 
of the flux quantum $\Phi_0 = h / e$. 
Therefore, the minimal non-zero magnetic flux through the system (the surface of the torus in Fig. \ref{torus}.b) is $\Phi_0$. Most numerical calculations use the Landau gauge with 
$A^{\mathrm{Landau}}_{(x,y),(x,y+1)} = \frac{x}{L} \, m$ with $m$ an integer
and $A^{\mathrm{Landau}}_{(x,y),(x+1,y)} = 0$, which results in a total flux
\be
\Phi_{\mathrm{Landau}} = m \cdot L \, \Phi_0
\ee
through the system. The minimal non-zero flux in the Landau gauge is thus $L$ times larger than the flux quantum $\Phi_0$, and, consequently, the possible values of magnetic field,
$B_{\mathrm{Landau}} = \frac{\Phi_{\mathrm{Landau}}}{L^2} = \frac{m}{L} \, \Phi_0$, are restricted and rather large.

Clearly, to perform efficient finite size scaling at a fixed magnetic field, one needs to construct a lattice gauge, which is able to produce magnetic fields below the Landau
gauge limit, $B^{\min}_{\mathrm{Landau}} = \frac{1}{L} \Phi_0$.
Here we propose to use a lattice gauge, as illustrated in Fig. \ref{gauge}, that realizes the minimal flux and the corresponding minimal magnetic field. Along the $y$ bonds,
we use a Landau gauge
\be
A_{(x,y)(x,y+1)} = m \frac{x}{L^2} \; \phantom{ab} x \in {1\dots L} \mathrm{.}
\ee
This is by a factor $1/L$ smaller than the usual Landau gauge and, consequently, amounts in an additional jump in the phase of the hopping between lattice sites $x=L$ and $x = 1$,
$\Delta \varphi = - 2 \pi \, m \frac{1}{L}$.
Such a jump would introduce a strong magnetic field at the boundaries, if it is not compensated.
Therefore at the boundary between $x = L$ and $x = 1$, we apply a lattice vector potential in the $x$ direction\cite{footnote2} 
\be
A_{(x=L,y)(x=1,y)} = - m \frac{y}{L} \; \mathrm{.}
\ee
One can verify that the magnetic field in each cell is $\Phi_0 / L^2$, therefore the total magnetic flux is just the minimal non-zero flux $\Phi_0$. Within this new gauge,
we can thus reach magnetic field values of $B = \frac{m}{L^2} \, \Phi_0$. This new gauge allows us to change the system size in relatively small steps when the magnetic field is fixed.

\section{Results}
\subsection{RG flow and critical behavior}
\begin{figure}
 \includegraphics[width = 0.4 \textwidth]{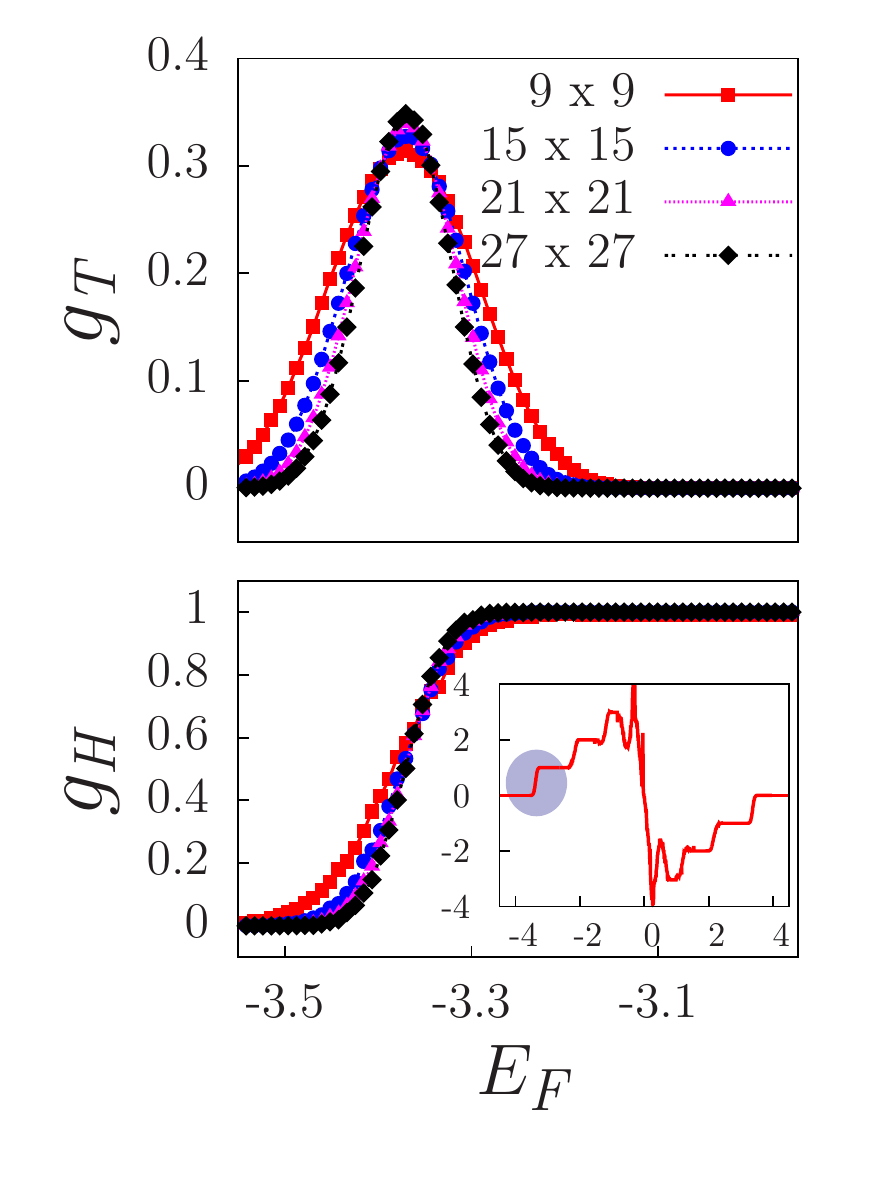}
 \caption{\textit{(Color online)} Thouless conductance (upper panel) and Hall conductance (lower panel) as a function of Fermi energy around the first Landau band for 
 $B = \Phi_0 / 9$ and  $W = 1$. 
System sizes are $L = 9 \, , \;  12  \, , \; 18 \, , \; 27$. \textit{Inset:} Hall conductance $g_H$ as a function of the Fermi energy, $E_F$, in the whole band. 
The shaded region 
highlights the first QH step, shown in the main panel.
In the lower half of the band electron-like behavior, while
in the upper half hole-like behavior is observed.}\label{ConductancesEF}    
\end{figure}

Let us start by analyzing the critical behavior of the dimensionless conductances.
The Thouless  and Hall conductances were calculated for system sizes between $L=9$ and $L=33$, magnetic fields  $B = 1 / 9$, $1 / 4$ and $1 / 16$ (in units of $\Phi_0 / a^2$) , 
and disorder 
strengths  $1 \le W \le 3.5$. A total number of $\sim 5 \cdot 10^8$ eigenstates were computed for each system size, magnetic field, and disorder. 
The behavior of the ensemble averaged conductances as a function of the Fermi energy is shown in Fig. \ref{ConductancesEF}.
The Hall step becomes sharper with increasing system size and the peak in the Thouless conductance gets sharper as well.

Based upon the two parameter scaling theory, near the transition, the dimensionless Hall conductance is expected to scale with the system size as
\be\label{gHfit}
g_H(L') - g_H^* \cong \left(\frac{L'}{L}\right)^{y}(g_H(L) - g_H^*) \; \textrm{,}
\ee
where $y$ is the scaling dimension of the Hall conductance, and $g_H^*$ denotes the critical Hall conductance. 
In contrast, the Thouless conductance is predicted to be an irrelevant scaling variable on the critical surface, where
\be\label{gTfit}
g_T(L') - g_T^* \cong \left(\frac{L'}{L}\right)^{-|y_2|}(g_T(L) - g_T^*) \; \textrm{,}
\ee
with $y_2$  the scaling dimension of the leading  irrelevant operator.
 We estimated the critical values of the Hall and Thouless conductances and the exponents $y$ and $y_2$ by performing a finite size scaling analysis, yielding:
\be
g_H^* = 0.612 \pm 0.023 \phantom{a}, \phantom{bc} y = 0.351 \pm 0.082 \; , 
\ee
and
\be
g_T^* = 0.386 \pm 0.011 \phantom{a} , \phantom{bc} |y_2| = 0.43 \pm 0.14 \; \textrm{.} 
\ee
The critical exponents $y$ and $y_2$ agree within our numerical accuracy with the values $y = 1/\nu \approx 0.385$ and $|y_2| \approx 0.4$,  extracted 
through transfer matrix methods.\cite{QHallExponent_Slevin,QHallIrrelevant,Evers}
\begin{figure}
 \includegraphics[width = 0.48 \textwidth]{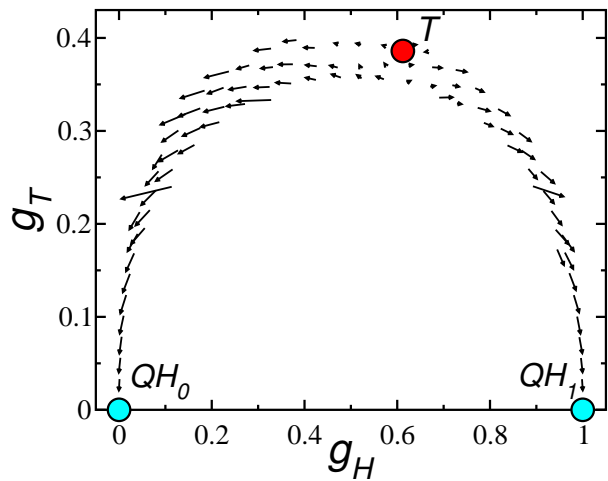}
 \caption{\textit{(Color online)} Two parameter renormalization flow extracted from finite size scaling for $B = \Phi_0 / 4$, $W \in 2 \dots 3.3$ and
 $L \in 12 \dots 24$.
 The arrows show the direction of increasing system size. The extrapolated position of the critical point is denoted by a red circle ($T$), the zeroth and first QH fixed points are
 denoted by cyan circle ($QH_0$ and $QH_1$).}\label{flow}
\end{figure}

\begin{figure}
 \includegraphics[width = 0.42 \textwidth]{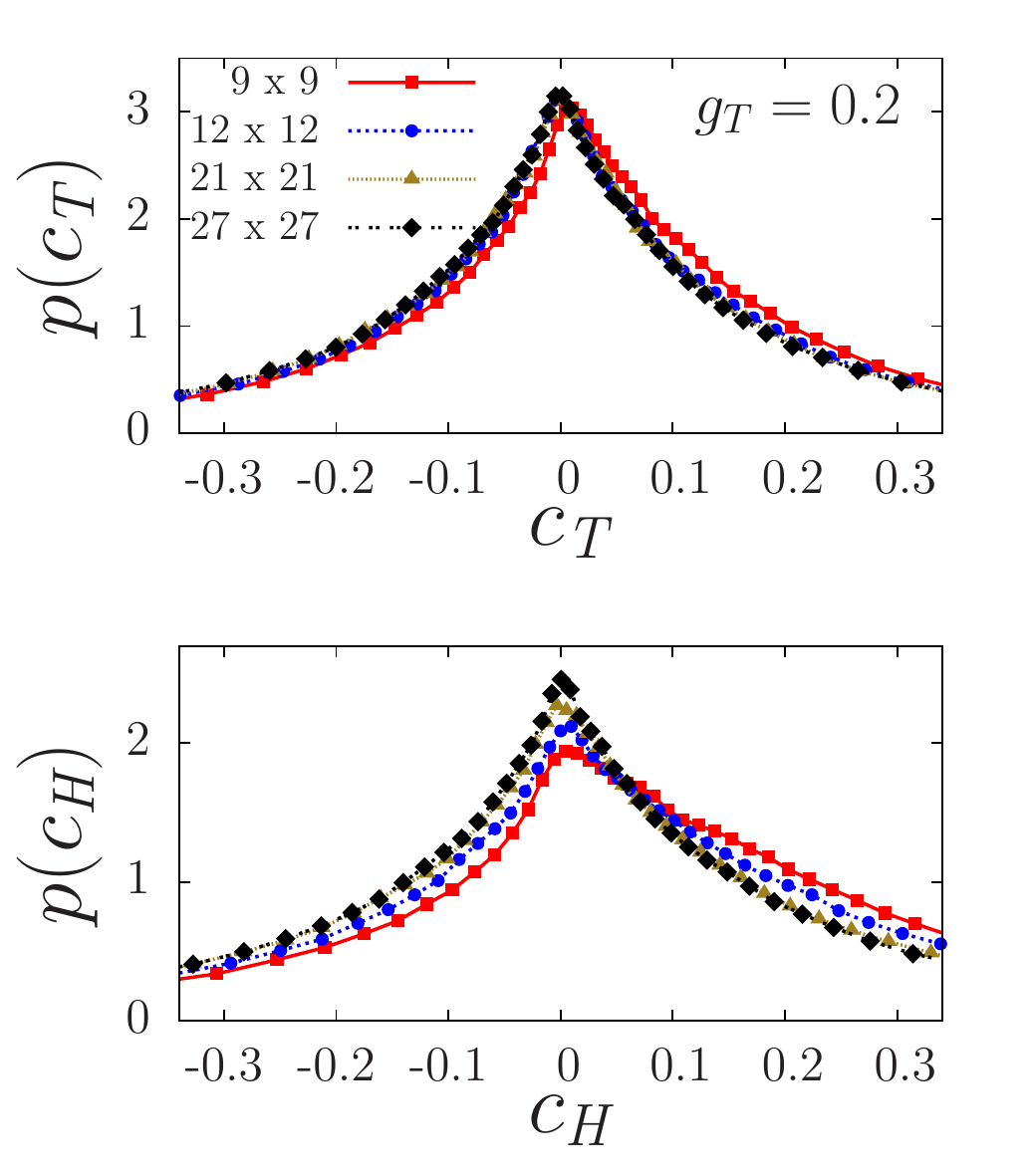}
 \caption{\textit{(Color online)} Distribution of the Thouless curvature (upper panel) and the level resolved Hall conductance (lower panel). 
 We used B=1/9 and W=1, and varied the system size (see legend). 
 For each system size we selected energy regions corresponding to a fixed $g_T= 0.2$, and determined the distributions for a large number of disorder realizations. 
 Distributions for a fixed $g_T$ depend explicitly on the system size, $L$, but converge to a limiting distribution for large $L$ (see data for $ L = 21$ and $27$).}\label{pcTgH}
\end{figure}
The system size driven $(g_T,g_H)$ flow is displayed in Fig. \ref{flow}. The qualitative agreement with the Pruisken-Khmelnitskii scaling is apparent.
As mentioned before, the new gauge introduced above enables us to increase the system size in smaller steps, and to get a much better resolution. Nevertheless, it remains challenging to collect
data from the exterior or deep interior of the critical dome (flipped ``U'' shape), because the trajectories remain always close to it. 
Interestingly, the flow is slightly asymmetrical, and the critical point is closer to the $n=1$ QH state than the trivial n=0 state. 
We do not have a firm explanation for this asymmetry. The lack of electron-hole symmetry could provide a natural explanation of such asymmetry. 
However, the fact that the flows extracted for various fillings overlap within our numerical accuracy, seems to rule out this possibility.
The observed asymmetry may also be a peculiarity of lattice calculations or non-universal finite size corrections. 

\subsection{Curvature distributions}
The presence of two characteristic scaling variables is also clear from a careful analysis of the distribution of level resolved Hall conductances and Thouless curvatures. 
Single parameter scaling\cite{GangOfFour} would imply that 
these distributions should be characterized by a single dimensionless parameter, which we can choose to be the Thouless
conductance, $g_T = g_T(W,E,L,B,\dots)$. To test the single parameter scaling hypothesis, we selected regions in the energy spectra with a fixed 
Thouless conductance, $g_T$ (i.e., fixed average absolute curvature $\overline{| c_T |}$), and determined the distributions $p(c_T | g_T, L, W, B)$ and
$p(c_H | g_T, L, W, B)$.\cite{footnote3} We found that the single parameter scaling hypothesis is clearly violated for small system sizes; 
both $p(c_T)$ and $p(c_H)$ depend explicitly on the system size, $L$. The explicit $L$-dependence is more pronounced in the distribution of the level resolved Hall conductance, 
but can also be seen
in the distribution of the level curvatures. Increasing L, however, the distributions converge to a limiting distribution (see data for $L = 21\; \textrm{and} \; 27$ in Fig. \ref{pcTgH}). 
This behavior can be understood in terms of the two parameter scaling theory. According to the latter, the distributions $p(c_T)$ and $p(c_H)$ depend on two dimensionless parameters,
$g_T$ and $g_H$: $p(c_T) = p(c_T | g_T, g_H)$ and $p(c_H) = p(c_H | g_T, g_H)$. For a given value of $g_T$, increasing $L$ moves the corresponding $(g_T, g_H)$  point towards the 
flipped ``U'' envelope in the $(g_T, g_H)$ plane. That means that for systems with large $L$, $g_H$ becomes effectively a function of $g_T$, $g_H \rightarrow g_H(g_T)$, and therefore 
$p(c_T)$ depends solely on $g_T$.

\begin{figure}
 \includegraphics[width = 0.42 \textwidth]{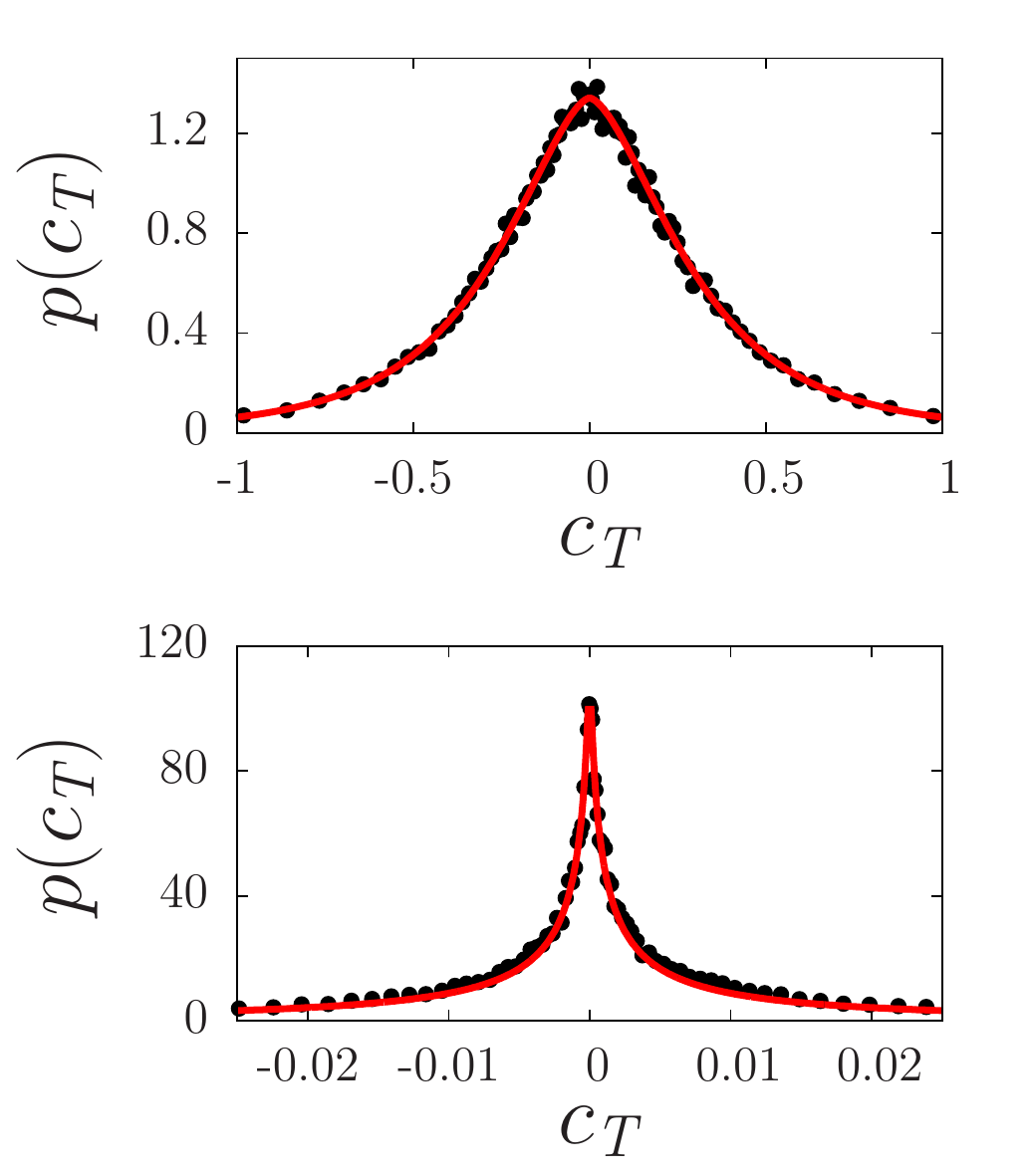}
 \caption{\textit{(Color online)} Curvature distributions in the vicinity of the transition fixed point (upper panel, $g_T \approx 0.35, g_H \approx 0.55$),
 and close to the Quantum Hall fixed points (lower panel, $g_T \approx 0.003$, $g_H \approx 0.001$ ). Continuous lines denote  modified Cauchy (top) and lognormal (bottom) fits.}\label{pcTfit}
\end{figure}
The distributions $p(c_T)$ and $p(c_H)$ vary considerably within the $(g_T, g_H)$ plane (see Fig. \ref{pcTfit}). Near the transition point, the distribution of the dimensionless 
Thouless curvatures can be 
well fitted by a modified Cauchy distribution,
\be
p(c_T) \propto \frac{1}{(c_T^\kappa+a)^{(2+\beta)/\kappa}} \; \textrm{,}
\label{curv_distr}
\ee
with the constant $\beta=2$ characterizing the  unitary ensemble, and   $\kappa$ a
symmetry class dependent anomalous dimension.
Such a distribution has been conjectured for the critical curvature distribution in orthogonal and unitary ensembles, and verified 
numerically for the orthogonal case.\cite{Kravtsov,footnote4} By fitting
the numerically obtained distributions, we extract an exponent
\be 
\kappa = 1.603 \pm 0.026 \; .
\ee 
This value is close to the exponent $\kappa = 2$, predicted for disordered metallic systems in the unitary ensemble by random matrix theory.\cite{vonOppen_GUE,vonOppen_GUE2} In fact, although a modified
Cauchy distribution is needed to reach a high quality fit of the small curvature part of the distribution, the random matrix expression ($\kappa = 2$) also provides an acceptable
fit 
of the data.

Close to the attractive Quantum Hall fixed points, on the other hand, the dimensionless curvature is lognormally distributed with a good accuracy, a behavior characteristic of strongly
localized states.\cite{Titov}

\section{Conclusion}
In this work, we investigated disordered Quantum Hall systems by performing numerical computations within a torus geometry. 
We introduced a new magnetic gauge, which enabled us to reach the smallest magnetic field allowed by the periodic boundary condition, 
$B = \frac{1}{L^2} \frac{h}{e}$. With this
new gauge, we were able to increase the system size in smaller steps, and could perform efficient finite size scaling.

We determined the boundary condition (phase) dependence of the eigenstates and eigenenergies, and computed from these the diagonal and Hall conductances.
We established the system size driven renormalization flow of the dimensionless conductances, and found it to be consistent with the theoretical predictions of Pruisken and Khmelnitskii.
We identified the Quantum Hall
fixed points, responsible for the quantized values of the Hall conductance, and the critical fixed point characterizing
the transition between neighboring Quantum Hall phases. In the vicinity of this critical point, the Hall conductance is found to be a relevant scaling variable,
while the diagonal conductance becomes irrelevant. We estimated the critical exponents of the 
transition fixed point, and found them to agree with the values calculated using transfer matrix methods. 

We investigated the distributions of level curvatures, and observed a clear violation of the one parameter scaling, demonstrating the necessity of a second parameter.
For large system sizes, however, the system flows towards a critical line, and the single parameter scaling is found to be restored, 
in agreement with the Pruisken-Khmelnitskii scaling theory. 
Near the critical point, the distribution of the Thouless curvature is found to agree with the predictions of random matrix theory (Gaussian Unitary Ensemble). 
Close to the Quantum Hall points the curvature distribution is lognormal.

This research has been supported by the
Hungarian Scientific Research Fund OTKA under Grant Nos.
K105149, and  CNK80991. We also acknowledge the support of the Helmholtz Virtual Institute "New states of matter and their excitations" as well as 
the DFG Schwerpunkt 1666 Topological Insulators, and a Mercator Guest Professorship.


\begin{thebibliography}{99}
\bibitem{footnote1}The Hall conductance can have nonquantized values however if the localization length at the Fermi energy is larger than
other length scales.
\end{thebibliography}
\end{document}